\documentclass[10pt,notitlepage,nofootinbib,aps,pre,floats,onecolumn]{revtex4}

\usepackage{graphicx}
\usepackage{graphics}
\usepackage{amsmath}
\usepackage{amsfonts}
\usepackage{amssymb}
\usepackage{epstopdf}
\usepackage{makeidx}
\usepackage{epsfig}
\usepackage{bm}
\usepackage{times}
\usepackage{hyperref}
\usepackage{color,soul} 
\usepackage[english]{babel}
\usepackage{threeparttable}
\usepackage{booktabs}
\usepackage{array}
\usepackage{multirow}
\usepackage[symbol]{footmisc}

\AtBeginDocument{ 			
    \heavyrulewidth=.08em
    \lightrulewidth=.05em
    \cmidrulewidth=.03em
    \belowrulesep=.65ex
    \belowbottomsep=0pt
    \aboverulesep=.4ex
    \abovetopsep=0pt
    \cmidrulesep=\doublerulesep
    \cmidrulekern=.5em
    \defaultaddspace=.5em
}

\hypersetup{
    bookmarks=true,         
    unicode=false,          
    pdftoolbar=true,        
    pdfmenubar=true,        
    pdffitwindow=false,     
    pdfstartview={FitH},    
    pdfauthor={Vicens, J},  
    colorlinks=true,        
    linkcolor=black,         
    citecolor=black,        
}

\graphicspath{{}} 

\definecolor{dkgreen}{rgb}{0,0.6,0}		
\definecolor{gray}{rgb}{0.5,0.5,0.5}	
\definecolor{mauve}{rgb}{0.58,0,0.82}	

\renewcommand\theequation{\arabic{equation}}
\renewcommand\thefigure{\arabic{figure}}
\renewcommand\thetable{\arabic{table}}

\newcolumntype{C}[1]{>{\centering\arraybackslash}p{#1}} 

\begin{document}

\title{Citizen Social Lab: A digital platform for human behaviour\\ experimentation within a citizen science framework}
\author{Juli\'an Vicens\textsuperscript{1,2,3}, Josep Perell\'o\textsuperscript{2,3} and Jordi Duch\textsuperscript{1}\footnote[1]{jordi.duch@urv.cat}}
\affiliation{\\}
\affiliation{\textsuperscript{1}Departament d'Enginyeria Inform\`atica i Matem\`atiques, Universitat Rovira i Virgili, 43007, Tarragona, Spain} \affiliation{\textsuperscript{2}Departament de F\'isica de la Mat\'eria Condensada, Universitat de Barcelona, 08028, Barcelona, Spain} \affiliation{\textsuperscript{3}Universitat de Barcelona Institute of Complex Systems UBICS, 08028, Barcelona, Spain}

\begin{abstract}
Cooperation is one of the behavioral traits that define human beings, however we are still trying to understand why humans cooperate. Behavioral experiments have been largely conducted to shed light into the mechanisms behind cooperation -- and other behavioral traits. However, most of these experiments have been conducted in laboratories with highly controlled experimental protocols but with limitations in terms of subject pool or decisions' context, which limits the reproducibility and the generalization of the results obtained. In an attempt to overcome these limitations, some experimental approaches have moved human behavior experimentation from laboratories to public spaces, where behaviors occur naturally, and have opened the participation to the general public within the citizen science framework. Given the open nature of these environments, it is critical to establish the appropriate data collection protocols to maintain the same data quality that one can obtain in the laboratories.  
In this article we introduce Citizen Social Lab, a software platform designed to be used in the wild using citizen science practices. The platform allows researchers to collect data in a more realistic context while maintaining the scientific rigour, and it is structured in a modular and scalable way so it can also be easily adapted for online or brick-and-mortar experimental laboratories. Following citizen science guidelines, the platform is designed to motivate a more general population into participation, but also to promote engaging and learning of the scientific research process. We also review the main results of the experiments performed using the platform up to now, and the set of games that each experiment includes. Finally, we evaluate some properties of the platform, such as the heterogeneity of the samples of the experiments and their satisfaction level, and the technical parameters that demonstrate the robustness of the platform and the quality of the data collected. \end{abstract}

\maketitle
\section*{Introduction}
Social dilemmas modeled as behavioral games are important tools to study the general principles of human behaviour and to understand social interactions. Social dilemmas occur when individual interests conflict with other individual or collective interests \cite{Schroeder1995, Kollock1998, Nowak2011}. Behavioural experimentation thus yield relevant scientific outcomes that have been used to test theories and to refine models, providing experimental data for simulations \cite{Sanchez2018}, and making the understanding of human behaviour move forward. But the impact of the experimental insights go beyond the scientific theories, because social dilemmas describe interactions and conflicts in real-life situations such as climate change mitigation, refugee repatriation, use of public space, social inclusion, gender discrimination, care-in community in mental health or resource depletion, and results obtained from behavioral research can be translated to improve all these areas.

Traditionally, most experiments have been conducted in laboratories with highly controlled experimental protocols but with limitations in terms of subject pool or decisions' context \cite{Levitt2007,Levitt2007a,Levitt2008,List2009}. There is a sample bias since a large number of laboratory experiments' sample consists of students who have a particular socioeconomic and sociodemographic situation. Thus, those studies do not reflect the general population behavior \cite{Levitt2007,Fehr2004a,Rosenthal1969,Orne1962}. Besides, generalizability of results of laboratory experiments also is affected by the physical context in which they are performed. The situations of social interaction that are studied do not happen in laboratories, but in real life scenarios where participants face dilemmas and make decisions. This leads participants in laboratories to not engage in real-world behaviors, but instead in behaviors that are biased by the experimental conditions. 

Furthermore, recently social experimentation has been affected by the general crisis of science in replicability and reproducibility, issues that concern the main actors in science \cite{Chang2015,OpenScienceCollaboration2015,Munfalo2017}. Some efforts have been done to solve this situation, promoting the transparency in the statistical and methodological aspects of laboratory work, but also promoting the publication of more detailed methods, the data sources and the codes used in the experiments and in the analysis \cite{Nature2017}.  Scientists are encouraged to conduct replication studies \cite{Editorial2017} and, in general, to pursue a more open research culture \cite{OpenScienceCollaboration2015,Nosek2015}. 

In recent years, Computational Social Science has emerged as a multidisciplinary field that studies complex social systems and provides new insights about social behaviour, combining tools and methods from social and computer sciences \cite{Lazer2009,Cioffi-Revilla2010,Conte2012,Mann2016}. In this line, a large number of studies have been conducted generally exploiting big amounts of social data, mostly collected from online social platforms (Twitter, Facebook, Coursera, etc.) \cite{Bond2012}.  Within the same field, some researchers have started to use online services such as Amazon Mechanical Turk as platforms to recruit and develop their behavioral experiments \cite{Mason012, Rand2012}. Many experiments have been successfully deployed in these services providing new insights to social problems from another perspective \cite{Shirado2017}, however experiments on this platform also suffer from some known limitations \cite{Stewart2017} 

Between the studies conducted with large-scale data from online platforms (that come from less controlled samples and protocols) and the small-scale data collected from the experimentation in behavioural science labs (collected with more robust protocols) there is a missing gap. New platforms fill this gap providing opportunities for the design of mid and large-scale behavioral experiments in online labs that guarantee the quality of the data collection \cite{Radford2016,Chen2016,Holt2005}. These more flexible platforms have great advantages, as (1) they facilitate the recruitment of more diverse sociodemographic profiles or from very specific communities according to the needs of the experiment, (2) they are able to carry out the experiments in a distributed way in space and time, and (3) they are more efficient at the economic level, since the infrastructure is much lighter. In these platforms other limitations arise, such as the identification of the experimental participants or the economic incentives, to mention only a few.


Our scenario of experimentation is described in the context of pop-up experiments \cite{Sagarra2015}, an intermediate situation between traditional behavioural experimentation and big data analysis. The basic idea is to translate the experiments outside the lab to real contexts, and to open participation to new and more diverse audiences. More importantly, the experiments are not only build by taking into account the researcher's interests and motivations, but also considering the perspective of citizen participation and its social impact in terms of providing the right knowledge to conduct new evidence-based policies by public administrations and empower participants to trigger civic actions. This is framed within the citizen science approach \cite{Bonney2014,Gura2013,Hand2010,Silvertown2009}, that promotes the participation and inclusion of non-expert audience in real research processes in different ways \cite{Kullenberg2016,Bonney2015} (co-creating projects, collecting data, interpreting and analyzing data, and provide actions based on the evidences collectively gathered). Citizen science helps us to involve the general public in behavioural experimentation and impacts the participants themselves \cite{Senabre2018,Bonney2015,Price2013,Bonney2009,Bonney2009a}, for instance increasing their disposition to science \cite{Perello2017}.

To carry out these experiments interactively, we designed and implemented Citizen Social Lab, a platform with a collection of decision-making and behavioral games based on a light infrastructure that can be installed and executed in real-life contexts in a simple but robust way. 
Depending on the goal of the experiment and the behavioral variables to be studied, the researcher can select and parametrize one or various games, and also define the general dynamics of each experimental session. The platform registers all the behavioural actions taken by the participants, but also provides surveys to collect sociodemographic data, information about the participants' experience or their decision making process. The platform does not allow the intervention of uncontrolled participants, and it registers data accurately without alterations of any kind.

In contrast to other existing platforms, this platform has been designed to follow citizen science guidelines and to be used in experimental settings where participants are recruited using opportunistic and random sampling. For these two reasons, both the experimental staging and the platform include features to attract the attention of participants and, once they are enrolled, to improve their focus and engagement within the experiment. In this line different approaches are used, one of them being the gamification of the experience \cite{Ponti2018,Bowser2013}, which consists in presenting the experiment as a game and a scientific investigation at the same time. Another important feature is the feedback and knowledge obtained by the participants after the experience, for instance through personalized reports for each participant or by organising public lectures that summarize the results once a paper has been published. These efforts also add new dimensions to the mandatory open data access or ethical and transparency requirements when dealing with citizen science approaches. 

The experimentation platform has been active since 2013 and, within that time, it has been used successfully in 15 experiments to study different aspects of human behavior. Up to this date 2821 people have contributed, taking around 45200 valid decisions. We have developed it as an open-source project where software developers and the scientific community can help grow the platform, but also to facilitate the reproducibility of the experiments and to foster the usage of platforms like these as an alternative method to conduct behavioral experiments in all types of environments and settings.

\section*{The platform}
\begin{figure}[!h]
\includegraphics[width=0.75\linewidth]{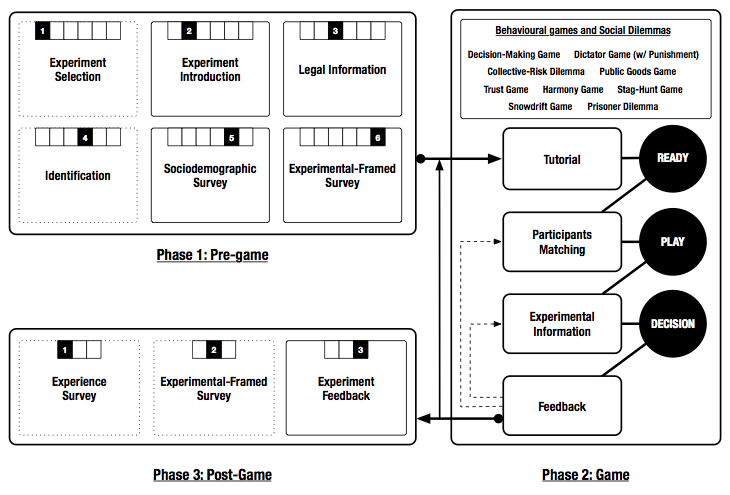}
\caption{Block diagram of a participant's flow through one experimental setup. The participant goes through three stages: the first stage contains the pre-game module with preliminary instructions about the experiment and surveys, the second stage contains the core game mechanics (which implements the suite of decision-making and behavioural games), and the third stage consists of the post-game module with the final feedback of the experiment and surveys about the experience and the topic of the experiment. Not all these modules and interfaces are present in all the experimental setups.}
\label{fig:client_platform}
\end{figure} 

Citizen Social Lab is a platform designed to assist in the deployment of human behavioural experiments. It has been created with three important goals in mind that foster versatility. First, the platform is based on light and portable technologies, so it can be used in open and diverse environments following the guidelines of popup experiments \cite{Sagarra2015}, but also in pure online or in more ``classical'' experimental laboratories. Second, it has been designed with a friendly user interface to facilitate participation to a broader population, and to engage and motivate participants to solve the tasks proposed in the experiment while they have an enjoyable experience. And third, it is structured in a way that it is easy to incorporate any type of social dilemma or behavioural game, as well as, any type of interaction: individual/computer, individual/individual and individual/collective.

The platform allows researchers to carry out a suite of dilemmas or behavioural games, which compose the core of the system. The system already contains a few different available dilemmas, which are described in the next section, and this number is expected to increase as new experiments are developed and deployed using the platform. Moreover, beyond the data collected from the participant's decisions,  the system is designed to collect complementary data (about sociodemographics, user experience or experiment-related questions) through surveys before or/and after the social experiment takes place. It also registers all the activity of the participants when using the platform, which can be used to infer other parameters (e.g. response time).

The platform architecture is highly modular and allows the researcher to construct personalized environments combining and parametrizing the modules they require for their particular experimental setting.  The basic client modules currently available are the following: 
(i) Introductory interfaces, with brief but detailed information about the topic and goals of the experiment and legal information with privacy policy; 
(ii) Questionnaires; that can be used to collect sociodemographic information and also to present specific questions related with the experiment topic or setting. Questionaries can be used before and/or after the main experiment.
(iii) Tutorial and instructions; so participants can learn the rules and the mechanics of the experiment by themselves (even though in the physical location there are always researchers to provide support if any question arises) and practice a few testing rounds of the game to familiarize themselves with the game interface.
(iv) Games and/or dilemmas; the core of the platform, the module that runs the experiment to collect the decisions of the participants. An experiment can incorporate only one game or a collection of them.
(v) Results; a set of interfaces designed to provide feedback to the participants on the outcome of their decisions in the experiment. This is crucial to increase the positive return that they obtain for participating in the experience.
Finally, (vi) the administration interface is composed by a set of pages that let the researcher to control the parameters of each session, monitor the evolution of a game, and overview the general performance during the experiment in real-time.

The modules are combined and configured to define what we call the participant's flow through the experiment (see Fig.\ref{fig:client_platform}). The system is designed to automatically guide the participants through all the stages without the need of interacting with a researcher (unless otherwise required by the participant), and it allows the existence of simultaneous games at different stages of the experiment. 

\subsection*{Games Module}

The main goal of the platform is to collect the decisions of the participants when they face different types of dilemmas that are analogies of real-life situations. Most of the dilemmas included up to now are social, which require synchronized interaction with other individuals, however the platform can also be used to study individual decision-making situations that do not require real-time interaction with other participants. 



The first social dilemma implemented is a generalized version of a simple dyadic game, where two people have to decide simultaneously which of the two actions they will select, and the outcome is the result of the combination of them. Depending on the values presented to the participants, the can face different types of games: a Prisoner's Dilemma \cite{Axelrod1981,Rapoport1965}, a Stag Hunt \cite{Skyrms2003}, a Hawk-Dove/Snowdrift \cite{Sugden2004,MaynardSmith1982,Rapoport1966} or a Harmony \cite{Licht1999}. These dilemmas can be used to measure two important features of social interaction, namely the temptation to free-ride and the risk associated with cooperation. 

The second social dilemma, the trust game (TG), or otherwise called the investment game, is used in order to measure trust and reciprocity in social interactions \cite{Berg1995}. In TG two players are given a quantity of money. The first player send an amount of money to the second player, the first player is informed that the money that he sends will be multiplied by a factor (e.g. three). The second player takes the action of give some amount of the multiplied money back to the first player, and then both receive their final outcome. 

The third social dilemma, the Dictator game (DG) can be used to measure rationally self-interest or distribution fairness \cite{Fehr2004}. In this game, the first player ``the dictator" splits an endowment between himself and the second player,``the recipient". Whatever amount the dictator offers to the second player is accepted, therefore the recipient is passive, cannot punish the dictator's decision. DG is not formally a game because the outcome only depends on the action of one player, in game theory those games are known as a degenerated game. However, there are a modified version of DG which includes a third player who observes the decision of the dictator and has the option to punish the dictator's choice. The third person receives an endowment that could choose to spend to punish the dictator, so that punishing has a cost for the punisher. 

The fourth social dilemma, is a variant of the public goods game, which is a collective experiment game in which the players with their contributions decide invest in public goods or keep their private goods. This particular version is known as collective-risk dilemma \cite{Tavoni2011,Milinski2008}, and consists of a group of people who must reach a common goal by making contributions from an initial endowment. If the goal is reached, every individual receives the part of the money not contributed. If not, a catastrophe occurs with certain probability, and all participants lose all the money they had kept.


The platform also includes a decision-making game, where participants have to make decisions having uncertain and/or incomplete information \cite{Gutierrez-Roig2016}. This game is played individually so there are no interactions with other players during the game. With this game we can study decision making strategies by controlling the type and amount of information that can be accessed by the participants.

All the dilemmas described previously can be parametrized to allow for different types of studies (for instance, controlling the values of the payoff matrix) or extended to include different variations when they are available. Also, starting from the implemented interaction structures (Fig.\ref{fig:interaction_types}), new dilemmas can also be constructed and added to the platform following a simple set of guidelines described within the code of the platform.

\subsection*{Participation and Motivations}




Moving the experiments out of the laboratories implies that usually the participants are not captive in advance, but instead opens the opportunity to attract new audiences from a broader population. The recruiting process in open environments -such as a games' festival or public spaces- is substantially different from the recruitment in laboratories, and is usually based on opportunistic sampling. This type of recruitment presents new challenges, since you have to attract the interest of the population through other types of incentives. In the pop-up experimental framework \cite{Sagarra2015} we usually include a narrative context and performative elements to capture the attention of the participants. However, once the attention of potential participants has been attracted, it is also even more important to present the experiment in a motivating way to guarantee their participation until the end of the session.

We use gamification techniques to the degree that the experimental settings allow us to ensure the scientific rigour of the experiments. Behavioral games and dilemmas per se already have elements and mechanisms of games such as: challenges, objectives, rules, reward, punishment, interaction, competition, collaboration, call-to-action, among others. Based on them, we create an experience where we present some of the experiments as games, with a narrative setting that creates a story surrounding the experiment. In some experiments, mainly the ones that took place within the DAU Festival, an actor is in charge of the recruitment characterized as the main character of the experiment (Mr. Banks or Dr. Brain, see Fig. S1 in the SI). 

The experiments are designed to enhance the motivations of the participants, not only from the perspective of games, but also to impact in the science disposition of participants, the understanding of science or the impact in social issues. This is the particular case of the framed experiments: The Climate Game, Games for Mental Health, the games for social change within the STEM4Youth project and the street art performance called urGentEstimar; all of them are focused on real social concerns: a collective climate action, the mental heath promotion of in-community care services or the concerns from several school groups related to social inclusion, use of public space and gender violence. Furthermore, beyond the economic incentive to participate (according to their performance in the game), participants also receive feedback on how their decisions and contributions could be translated into scientific research.

In our case, there are two types of participation according to the experiment context. Most of the experiments have been carried out in uncontrolled environments in terms of recruitment, without captive participants (e.g. festivals or public spaces). In specific cases, where the experiments were carried out in collaboration with local communities, the need to apply special recruitment techniques is not so important since the communities are usually involved in the design and the deployment of the experiment. In any case, to support the game-based approach, the platform allows the introduction of resources to include the narrative, always preserving the scientific rigour, and also provides features that can be used to create a gamified experience.

\subsection*{Technical Details}

\begin{figure}[!h]
\includegraphics[width=0.75\linewidth]{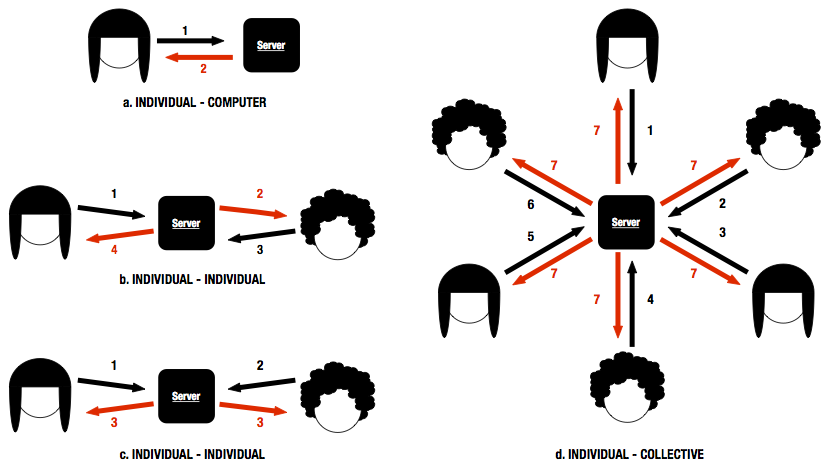}
\caption{Interaction types included in the platform. The platform currently implements four different types of interaction that cover individual-computer (a), individual-individual (b, c) and individual-collective (d) types of coordination. The numbers on the arrow indicate the order of when each interaction takes place, black arrows are interactions from individuals to the computer, and red arrows are interactions from the computer to the participants.}
\label{fig:interaction_types}
\end{figure} 

Some of the dilemmas previously explained require of individuals interacting in different manners. For instance, in games where two individual participate there are at least two possible interaction styles: one where the two individuals make a simultaneous decision without knowing the other's choice and after that they receive the outcome; or another where one player makes a decision while the other player is waiting, once the first decides the second, knowing the other's choice, makes her decision, finally both get their final feedback (see Fig. \ref{fig:interaction_types}). Also, experiments can have different evolution mechanics: from one-shot games, in which the players just make a unique decision, to iterated games in which the players make various decisions consecutively with the same or different participants. And finally, we also have to consider the possibility that the interaction between the players can be constrained by an underlying structure that defines the relationship between the players, which can range from a all-connected-to-all structure to a specific network structure.

Taking all these points into consideration, we designed a client-server architecture that controls the flow of the experiment according to the needs of the researchers. On one hand, the server manages the pace of the experiment, and implements all the core games and synchronization methods between players. It is based on a python-django backend, combined with a database to store the information generated separately by each experimental setup. The server can be run online, to allow experimentation on the internet or it can be installed in a local server to run experiments in local area networks. 

On the other hand, the client contains the user interface that the participants have to use to interact with the experiment. The technology on the client side is composed of html and javascript files that are generated dynamically from the experiment description files. The user interface has been designed to fit the resolution of a tablet device, but also works with any computer with a standard browser. It is also structured in a way that can be easily translated to other languages.

\begin{figure}[!h]
\includegraphics[width=0.75\linewidth]{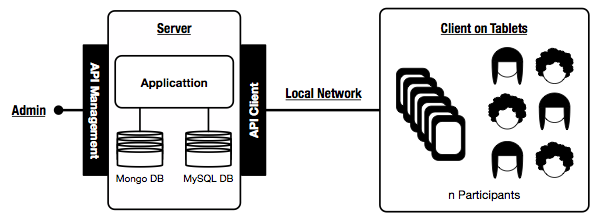}
\caption{Example of the platform infrastructure. This is the basic technological infrastructure used in the majority of experiments. It is designed to be rapidly deployed in any environment. }
\label{fig:infrastructure}
\end{figure}

Most of the experiments have used the same infrastructure consisting of a laptop that acted as a server and a collection of tablets that allowed up to 30 participants to be simultaneously participating in the experiment. In Fig. \ref{fig:infrastructure} we present a diagram of this infrastructure. Data is collected and stored in a database (which may be relational or not), and personal information is stored separately from the experimental data to follow the privacy guidelines required by this type of experiments.

Finally, live control of an experiment is critical to guarantee its correct development. For this reason, they can be controlled using an administration webpage that provides two features: it allows the researcher to configure the parameters that will be used in each iteration of the experiment (e.g. select if a certain group will be intervention or control) and it presents interfaces with the status of the experiment. Live monitoring can be done at two different scales, at a particular game level, where researchers have real-time detailed information about the evolution of a particular game (rounds played, decisions made, earnings, connection status, ...), or from a more general point of view to obtain a summary of the status of the experiment (demographics, games played, global earnings, ...).


%
%

\section*{The experiments}

\begin{table}
\centering
\setlength\extrarowheight{5pt}
\caption{Summary of experiments performed thus far. The suit of games is formed by: Decision-Making Game (DM), Harmony Game (HG), Snowdrift Game (SG), Stag-Hunt Game (SH), Prisoner's Dilemma (PD), Trust Game (TG), Dictator's Game (DG) and Collective-Risk Dilemma (CRD). The number of participants and decisions are the valid ones.}
\label{table:complete_experiments}
\resizebox{\columnwidth}{!}{
\begin{tabular}{cccccccccc}
\toprule
\multicolumn{1}{c}{\textbf{Experiment}} & 
\multicolumn{1}{c}{\textbf{Location}} & 
\multicolumn{1}{c}{\textbf{Date}} & 
\multicolumn{1}{c}{\textbf{Games}} & 
\multicolumn{1}{c}{\textbf{Participants}} &
\multicolumn{1}{c}{\textbf{Decisions}} &  
\multicolumn{1}{c}{\textbf{Publication}} & 
\multicolumn{1}{c}{\textbf{Data}}\\
\midrule
\multirow{3}{*}{\textbf{\shortstack{Mr. Banks:\\The Stock Market Game}}} 	& Barcelona & Dec.2013 & \multirow{3}{*}{DM}  	& 283	& 18525 &\multirow{3}{*}{\cite{Gutierrez-Roig2016}}  &\multirow{3}{*}{\cite{Gutierrez-Roig2016a}}\\
									& Brussels 	& Jul.2015 & 						& 37 	& 2397 	&											 &\\
									& Barcelona & Jun.2015 & 						& 20	& 1078 	&											 &\\ 
\midrule
\textbf{Dr. Brain} 					& Barcelona & Dec.2014 & HG, SG, SH and PD & 524  & 8366 &\multirow{1}{*}{\cite{Poncela-Casasnovas2016a}}  &\multirow{1}{*}{\cite{Poncela-Casasnovas2017}}\\ 
\midrule
\multirow{2}{*}{\textbf{\shortstack{Dr. Brain\\The Climate Game}}} 	& Barcelona & Dec.2015 & \multirow{2}{*}{CRD}  	& 320	&	3200	&\multirow{2}{*}{\cite{Vicens2017}}  &\multirow{2}{*}{(Embargoed)$^1$}\\
																	& Barcelona & Dec.2015 & 						& 100 	&	1000 	& 				 					&\\
\midrule
\multirow{4}{*}{\textbf{\shortstack{Games for\\Mental Health}}} 	& Lleida & Oct.2016 & \multirow{4}{*}{\shortstack{CRD, TG and PD}}	& 120 &1680 &	\multirow{4}{*}{\cite{Cigarini2018}}  &\multirow{4}{*}{\cite{Cigarini2018a}}\\
& Girona 	& Mar.2017 & & 60 	&	840 	& & &\\
& Sabadell 	& Mar.2017 & & 48 	&	672 	& & &\\
& Valls 	& Mar.2017 & & 42 	&	588 	& & &\\

\midrule
\multirow{3}{*}{\textbf{STEM4Youth}} & Badalona & Apr.2017 & \multirow{2}{*}{\shortstack{DG, TG and PD}}	 & 151	 & 1510	 &	\multirow{3}{*}{(In preparation)}  & \multirow{3}{*}{(Embargoed)$^1$}\\
& Barcelona & Sep.2017 & & 126	 & 1260	 & & & &\\
& Viladecans & May.2017 & \multirow{1}{*}{\shortstack{CRD}} & 162	 & 1620	& & & &\\

\midrule
\multirow{2}{*}{\textbf{urGentEstimar}} & T\`arrega & Sep.2017 & \multirow{2}{*}{\shortstack{DG, SG and PD}}	 & 756 & 2314 & \multirow{2}{*}{(In preparation)}  & \multirow{2}{*}{(Embargoed)$^1$}\\
& Barcelona & Oct.2017 & & 72 & 136 & & & &\\

\bottomrule
\end{tabular}
}
\\
\raggedright
$^1$ \small{Embargoed until scientific publication.}

\end{table}

The platform has been in use since December 2013 in 6 different experimental setups focused on the analysis of human behavior. Some of them have been repeated in different situations, which adds to a total of 15 experiments realized. In this section we describe the main goals and results of the six research projects based on this platform, which are also summarized in Table \ref{table:complete_experiments}.

\begin{enumerate}
\item The first experimental setup based on the platform is ``Mr. Banks: The Stock Market Game'' to study how people make decisions when they have limited and incomplete information. This setup emulated a stock market environment in which people had to decide whether the market would rise or fall. It allowed us to study the emerging strategies and the relevant use of information when making decisions under uncertainty, and the results are published in \cite{Gutierrez-Roig2016}. Three experiments based on this setup have been done in different locations, and is now available online\footnote[2]{http://www.mr-banks.com}.

\item Next, we created another experimental setup entitled ``Dr. Brain" to study the existence of cooperation phenotypes. The games played by the participants were based on a broad set of dyadic games and allowed us to deepen our understanding of human cooperation and to discover five different types of actors according to their behaviours \cite{Poncela-Casasnovas2016a}. 

\item The following experimental setup included in the platform was ``Dr. Brain: The Climate Game'', which was based on a collective-risk dilemma experiment to study the effect of unequality when participants face a common challenge \cite{Vicens2017}. Results showed that even though the collective goal was always achieved regardless of the heterogeneity of the initial capital distribution, the effort distribution was highly inequitable. Specifically, participants with fewer resources contributed significantly more (in relative terms) to the public goods than the richer - sometimes twice as much.

\item The fourth experimental setup implemented in the platform was called ``Games for Metal Health" which was repeated in 4 different locations. The goal of this project was to evaluate the importance of communities for effective mental health care by studying different behavioral traits of the different roles of the ecosystem. The results presented in \cite{Cigarini2018} reinforce the idea of community social capital, with caregivers and professionals playing a leading role. 

\item In the context of the EU project STEM4Youth we performed three experiments, which were co-designed with high-schools of Barcelona, Badalona and Viladecans. They addressed topics raised in workshops with students: gender inequalities, use of public space and integration of immigrants. The experiments combined a set of games that included Trust Game, Dictator's Game, Prisoner's Dilemma and Public Goods games.

\item Finally, we performed two experiments named ``urGentEstimar'' in the context of artistic performances in T\`arrega and Poblenou (a Barcelona neighborhood), in which the participants took part in a set of behavioural games which included Prisoner's Dilemma, Dictator's Game or Snowdrift, and which were framed around different concerns of local communities. 
\end{enumerate}


\section*{Platform evaluation}

In this section we analyze the versatility and the robustness of the platform by reviewing some of the results obtained by its use in different experimental setups. Mainly we focus on the sociodemographic diversity, the experience of participation, the time response data collected in the iterative experiments, and finally the robustness in the replicability of experiments. 

\subsection*{Sociodemogaphic}

\begin{figure}[!h]
\includegraphics[width=0.75\linewidth]{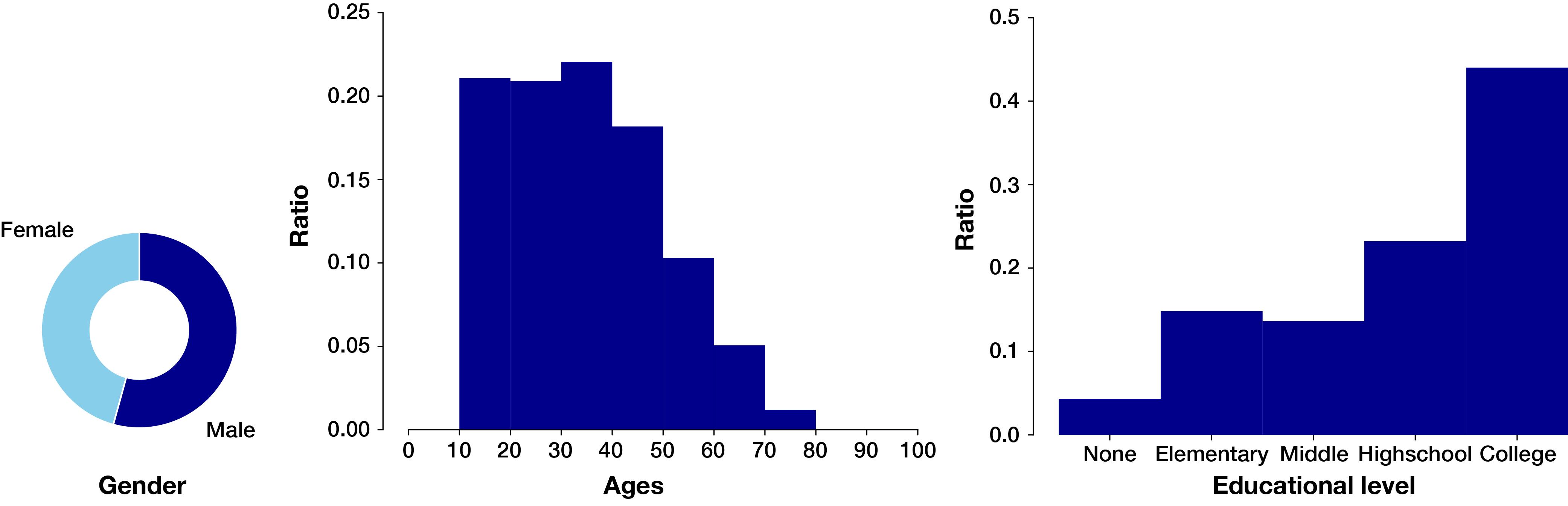}
\caption{Diversity of the participants pool. (Left) The proportion of participants in all the experiments (n=2821) regarding gender is 54.27\% males and 45.73\% females. (Center) Distribution of participants according to their ages in all the experiments (n=2821). (Right) Educational level of participants in all the experiments except ``urGentEstimar'', which didn't ask this question to participants (n=1993).}
\label{fig:diversity_participants_pool}
\end{figure}

To start, we review some of the demographical data of the participants in the different experiments. We already stated that one of the main goals of the platform was to open the experiments to a more general population. In this direction, in Fig.\ref{fig:diversity_participants_pool} we present an overview of the 2821 people that took part at some point in the behavioural experiments and perform the experiment with this plarform. We observe that we had a combination of participants from a wide range of ages, specially from 10 to 50, but older too, and diverse educational levels, with a predominance of those with higher education. Gender is also balanced (45.73\% females) compared with other similar experiments which are usually performed by students with sociodemographic bias.

\subsection*{Response times}

\begin{figure}[!h]
\includegraphics[width=0.75\linewidth]{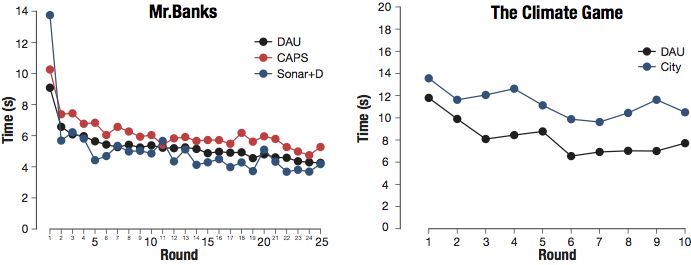}
\caption{Time of response in different games. (Left) Time response evolution across rounds in Mr. Banks experiments for the main performance in DAU (n=283) and the two replicas CAPS (n=37) and Sonar+D (n=20). (Right) Time response evolution across rounds in The Climate Game experiment in both performances, DAU (n=320) and City (n=100)}
\label{fig:times_reponse}
\end{figure}

The platform allows for the collection of very precise parameters about the participation in the experiments. One of them is the timestamp in which the participants perform an action. In iterated experiments, where participants make several decisions consecutively, the decision times are collected in each round so that we can calculate how long each participant takes to make a decision. An interesting parameter in behavioral experimentation is the learning time, or in other words, the evolution of time across the game.

In Fig.\ref{fig:times_reponse} we can see the evolution of the decision-making time across rounds. On the one hand, Mr. Banks presents the evolution of the three experiments that were carried out, the main one (DAU) and the two replicas (CAPS and Sonar+D). The evolution of the time response during the three experiments shows very similar trends. In the first round the time is substantially higher than the rest of the rounds and we see that from the 5th round the slope softens and stays more or less constant until the end. In this experiment, the variables that come into play to make a decision are the same round after round, so the trend is maintained during the game. The three experiments show similar trends but slightly different asymptotic values; the context, size and heterogeneity of the sample may be the cause of this variation, which confirms the accuracy of the data collected.

On the other hand, in the case of The Climate Game the evolution of the game is somewhat different. The game starts with long times that go down gradually; however, depending on the point of the game in which the participants are (i.e. the distance to the goal) the times increase or decrease. In this case, unlike the previous one, the decision at each moment is given by the distance to the final goal, so that, as they approach to the end of the game, the times increase again. Therefore, here we observe two sets of behavior: the learning at the beginning of the game and the uncertainty as the participants reach the last rounds. The trends of the two climate change experiments are similar, however, the absolute value of time is slightly higher in the ``City'' context.

\subsection*{Robustness of replicability}

\begin{figure}[h!]
\includegraphics[width=.7\linewidth]{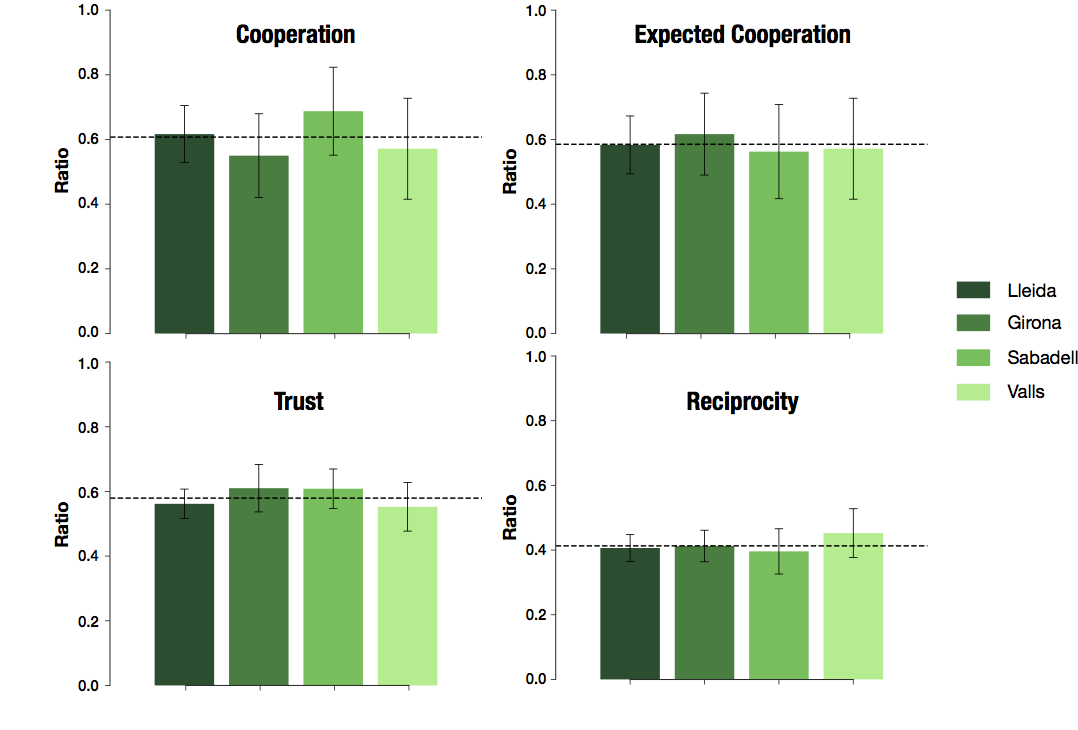}
\caption{Robustness of generalization in Mental Health experiments. Levels of cooperation, cooperation expectation, trust and reciprocity in the four experiments: Lleida (n=120), Girona (n=60), Sabadell (n=48) and Valls (n=42). It is represented the average level with 0.95 CI in each case. The dashed line represents the total average levels. There are no significant variation in the level of cooperation (Kruskal-Wallis, H= 2.38, p = 0.50), cooperation expectations (Kruskal-Wallis, H= 0.38, p = 0.94), trust (Kruskal-Wallis, H= 2.67, p= 0.45) and reciprocity (H= 3.02, p= 0.39). See Ref. \cite{Cigarini2018} for further details.}
\label{fig:mental_health_robustness}
\end{figure}

We also measure the consistency and the robustness of the results across different repetitions of the same experiment. Some of the six experimental settings described in the previous section were repeated in different environments and locations, in some cases with similar populations (e.g. the mental health experiment) and in other cases with different populations (e.g. the Mr. Banks experiment). We focus on Mental Health and Mr. Banks to examine the robustness on the platforms in order to collect quality data allowing the replicability in different situations.

Mental Health's experiments took place in Catalunya, in four different locations and social events (popular lunch, snack, etc.), in sum participated around 270 people. We analize the differences between the four events in cooperation, expected cooperation (Prisoner's Dilemma) and, trust and reciprocity (Trust game). The differences among the experiments in the four locations are not significative and the data can be aggregated to be analized as a whole \ref{fig:mental_health_robustness}.

\begin{figure}[h!]
\includegraphics[width=.7\linewidth]{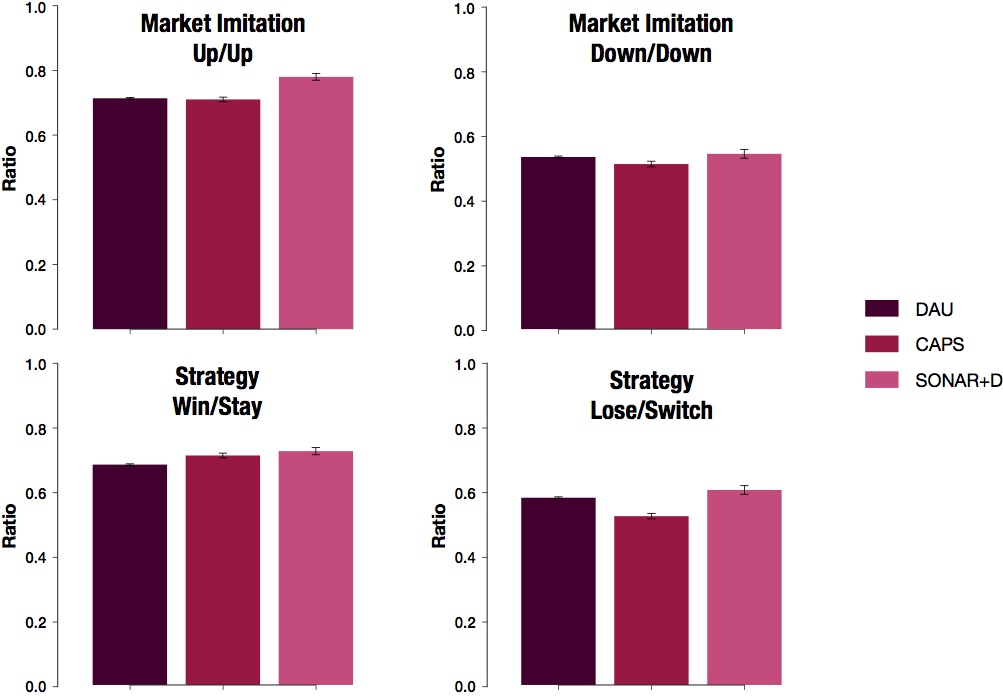}
\caption{Stability of strategies in Mr. Banks replication experiments. Ratio to follow strategies of Market Imitation and Win-Stay Lose-Shift in the experiments: DAU (n=283), CAPS (n=37) and Sonar+D (n=20). There are no significant differences in Market Imitation strategies except the probability to Up/Up  between the experiments of DAU and Sonar+D in (-2.53 SD). There are no significant differences in Win-Stay  except in the last case (Lose-Switch) between the experiments of DAU and CAPS (2.35 SD). See Table S1 and Table S2 for further details.} 
\label{fig:mr_banks_robustness}
\end{figure} 

Mr. Banks' experiment was performed in a main location, the DAU Festival, with a large participation, 306 people (283 valid participants), and obtaining robust results. From the analysis of decision emerged two strategies Market-Imitation and Stay-Win Switch-Lose. We compare the main result with two replicas that took place in two different events in Brussels (CAPS conference) and Barcelona (Sonar+D) with data from a narrow demographic populations and with the number of samples much lower than the main experiment. There are no significant differences ($>$1.96 SD) between the main experiment and the replicas except in Market-Imitation Up/Up between DAU and SONAR+D and Lose-Switch strategy between DAU and CAPS as Fig.\ref{fig:mr_banks_robustness} shows. This means that the platform captures data accurately since we are able to observe that the behavioural patterns found are consistent with the main results, because part of the results arise significant differences and the rest do not depend on the conditions.

\subsection*{Experience}

\begin{figure}[h!]
\includegraphics[width=.3\linewidth]{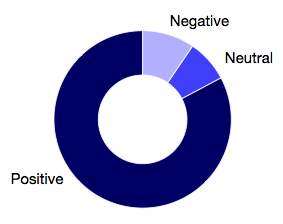}
\caption{Participants experience. Experience of participation in Mr. Banks, Dr. Brain and The Climate Change (n=1178). The most of participants (82.77\%) had a positive experience and a small group (9.51\%) had a negative experience, the rest (7.72\%) has an indifferent experience.}
\label{fig:participant_experience}
\end{figure}

Finally, another important aspect that we measured is the overall satisfaction of the participants after they finish the experiment. In the post-game survey of three games (Mr. Banks, Dr. Brain and The Climate Game) we asked the participants their level of satisfaction of the overall experience. Results of this question are presented in Fig.\ref{fig:participant_experience}. In all the experiments participants were mostly very satisfied or satisfied after the experience, specifically 82.77\%. The complete set of results about the experience in each game is represented in the Table S3. 

\section*{Discussion}

With Citizen Social Lab we present a platform that combines human behavioral experiments with a citizen science approach with the sake of bringing science to a broader audience and to perform social experiments beyond the laboratories. The platform is designed to be versatile, easy-to-use and robust, and to be used in open and diverse environments. It has already been adopted in several experiments by thousands of participants from a wide range of demographics, which mostly valued to experience to be very positive. The results obtained by some of the 15 experiments realized with the platform have also shown the scientific validity of the data obtained from the platform with several scientific contributions \cite{Gutierrez-Roig2016,Poncela-Casasnovas2016a,Vicens2017,Cigarini2018}. 

In order to maximize participation and make it much more diverse than usual social experiments, we move the laboratory to the wild. In this non-friendly context we use the the pop-up experimental setup to draw the attention of the potential participants (which are all the people of the surroundings) with different techniques described in Ref. \cite{Sagarra2015}. Then, we benefit from the lure of the game-base mechanics included in the platform in order to introduce them in the experience and guide them through all the tasks required by the experimental setup. This approach has proved to be very particularly successful in environments where they are likely to play (as the case of a games festival), leading to successful experiments with a high participation. 

In this line, it is important to emphasize the need to adapt the experiments to the environment where they take place (e.g. musical festivals, scientific conferences, and so on), especially the experimental design and the interaction with the platform, because it is the way to increase the empathy with potential participants. To achieve this, all the mechanisms of behavioral games and social dilemmas can be used to convert the interaction with the platform into a game (or other mechanism that fits in the context), always with the constraints imposed by the experimental scientific rigor. It is also important to remark that the interface of the platform has to be friendly and adapted to the latest usability standards to overcome the ``technological barrier'' that might appear for certain groups of ages or social backgrounds. For instance, in the experiments where kids are involved (which have been approved and designed accordingly), a friendly and visual appealing interface based on tablets provides an extra motivation to attract them to participate and reduces the time they need to learn the basics of the experiment.

After all the experiments and their repetitions we consider that the platform has already reached a high maturity level, but there are several points that still need some work to keep improving the technical and experimental parts. First, the platform has been largely tested within the pop-up experimental setting in physical environments. However, even when it has been designed to be easily integrated with online recruiting systems (e.g. Amazon Mechanical Turk), it has not been properly tested and validated in these environments. There is an opportunity to repeat some of the experiments to extend the consistency of the results when the dilemmas are presented to a purely online community.

Moreover, the platform is also constantly improving to provide new features and social dilemmas for the researchers. For example, we are creating the capacity for participants to create a unique profile and join in different environments. The long-term goal is to create a community of volunteers that participate in the experiments, and that can receive alerts when new opportunities to participate are open. We are also extending the number of available dilemmas within the platform as new research projects emerge which, once programmed and tested, are included in the main collection of available dilemmas.

The conceptual design in both types of experiments, the pop-up ones that have been done so far and the large-scale ones that are planned in the future, have in common that the motivations of participants and scientific rigor are at the center of the participatory design. The platform has room for improvement in motivating the participants and in offering rewards at the level of learning and participation. On one hand, it is necessary to improve the mechanisms of learning about the scientific topic of experimentation during the participation in the experiment, but also about the nature of their contributions and about the positive impact in carrying scientific knowledge forward. In this sense, many experiments are framed within a context of social impact, so participation can also be associated to a call to action to solve social concerns. In the most recent experiments, this type of actions have been carried out outside the context of the platform, however, the online version can also contribute to this mission.
 
On the other hand, participants can improve their experience at the end of the experiment, not only receiving the necessary economic incentive but also obtaining an on-site feedback expanded with real-time information about the research process in which they have participated. They can also obtain an improved experience by remotely following the evolution of the scientific research and participating in more phases of the scientific process. Another possible avenue to improve the platform is to build effective and real-time tools attached to experiments. Participants could in this way provide more feedback and actively contribute in the data interpretation and knowledge building process in both individual and aggregated levels. This effort appears to be meaningful to increase the participants' sense of ownership of the knowledge being produced by means of citizen science strategies.

Finally yet importantly, all the experiments done within these platform have been following open principles: the articles have been published as open access, and the data generated in all the experiments is also available in public repositories (properly anonymized) \cite{Gutierrez-Roig2016a, Poncela-Casasnovas2017, Cigarini2018a}. In the same vein, we are releasing the source code of Citizen Social Lab, including the core of the platform and the code of all the experiments done up-to-date, to the researcher community so they can use it to create their experiments using the templates and guidelines already established in the platform. The project code is going to be released under a CC BY-NC-SA license. In the very end, if we aim to practice citizen science, it is also necessary to claim for opening the platform by all means: releasing data and code and opening up the results to make them accessible and understandable for anyone. 

\section*{Acknowledgments}
%
We acknowledge the participation of 2821 anonymous volunteers who made the experiments based on the platform possible.  
We are grateful to N. Bueno, A. Cigarini, J. Gomez-Garde\~nes, C. Gracia-L\'azaro, M. Guti\'errez-Roig,  J. Poncela-Casasnovas, Y. Moreno, and A. S\'anchez for his work on the experiments and for useful discussions and comments about the platform and this article. We also thank the support of Mensula Studio for providing the graphical design for the experiments.
This work was partially supported by MINECO (Spain) through grants FIS2013-47532-C3-1-P (JD), FIS2016-78904-C3-1-P (JD), FIS2013-47532-C3-2-P (JP), FIS2016-78904-C3-2-P (JP), by Generalitat de Catalunya (Spain) through Complexity Lab Barcelona contracts no. 2014 SGR 608 and no. 2016 SGR 1064, (JP) and through Secretaria d'Universitats i Recerca contract no. 2013 DI 49, (JD, JV), by European Union Horizon 2020 research and innovation project STEMForYouth grant agreement no 7010577, (JV and JP).

\newpage
\appendix

\renewcommand\theequation{S\arabic{equation}}
\renewcommand\thefigure{S\arabic{figure}}
\renewcommand\thetable{S\arabic{table}}

\title{Supplementary Information}
\section{Supplementary Notes}

\subsection*{Platform screenshots}

\begin{figure*}[!th]  
\begin{center}$
\begin{array}{ll}
(a) \includegraphics[width=0.45\textwidth]{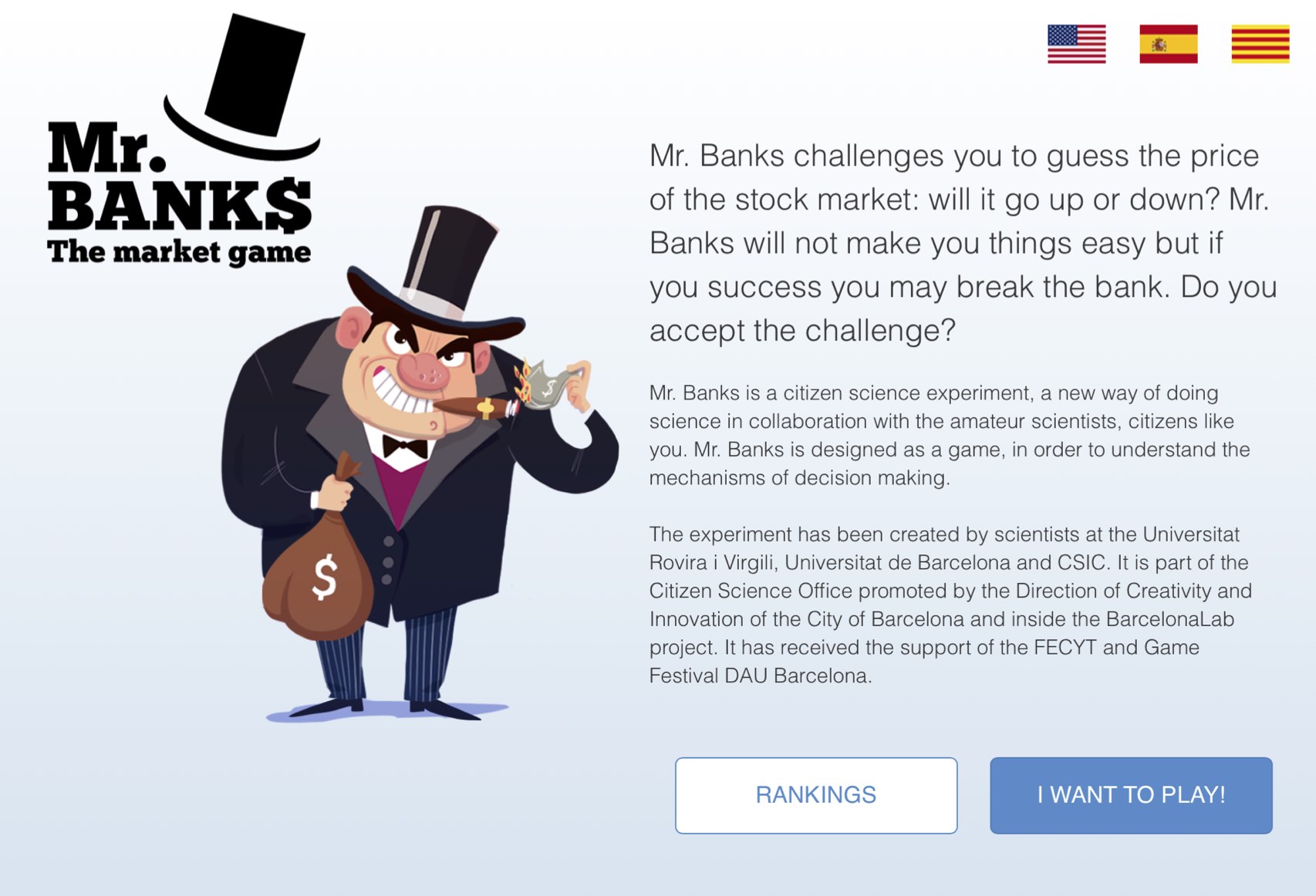}

\vspace{2mm} 

(b)
\includegraphics[width=0.45\textwidth]{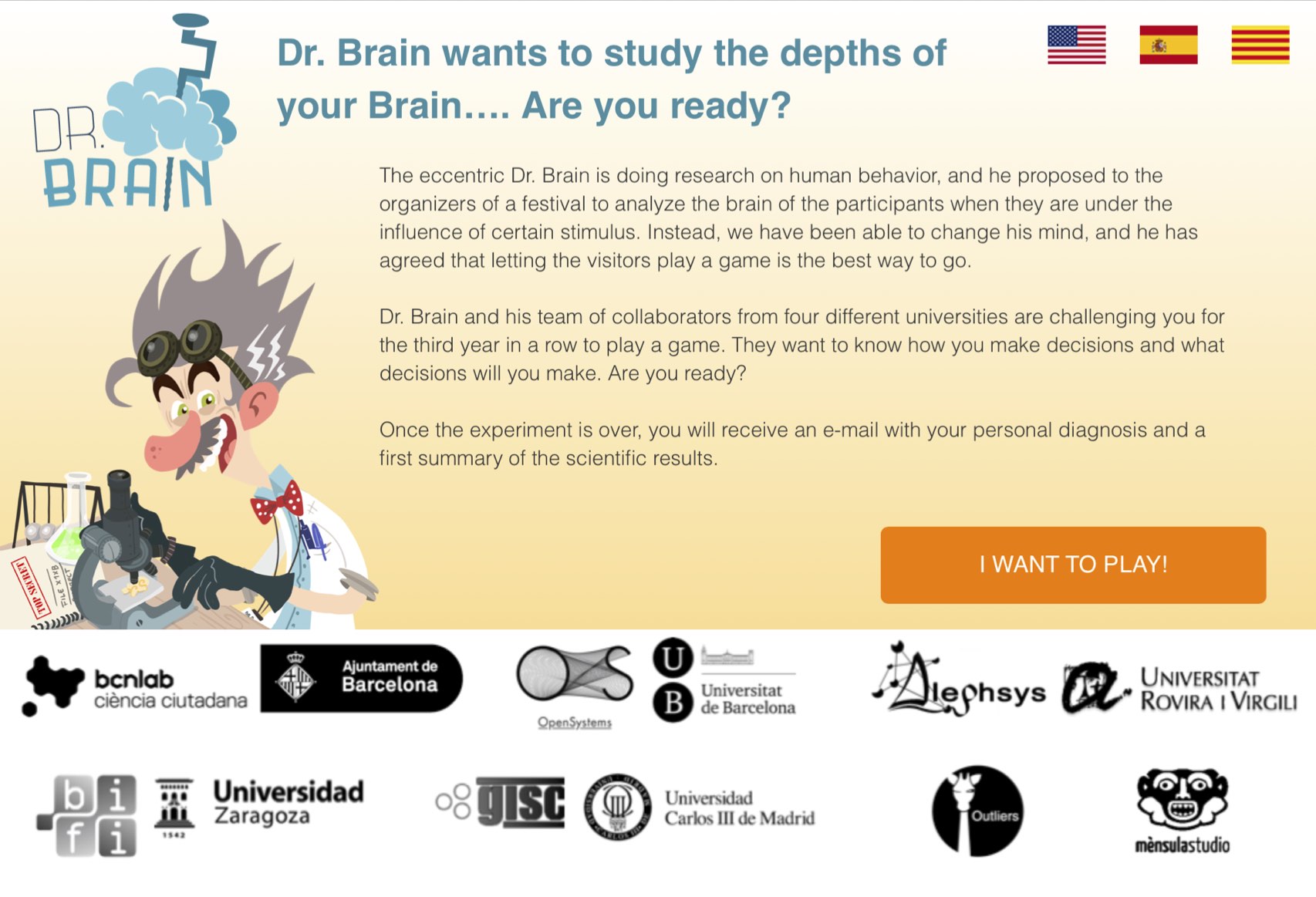}\\

\vspace{2mm}

(c) 
\includegraphics[width=0.45\textwidth]{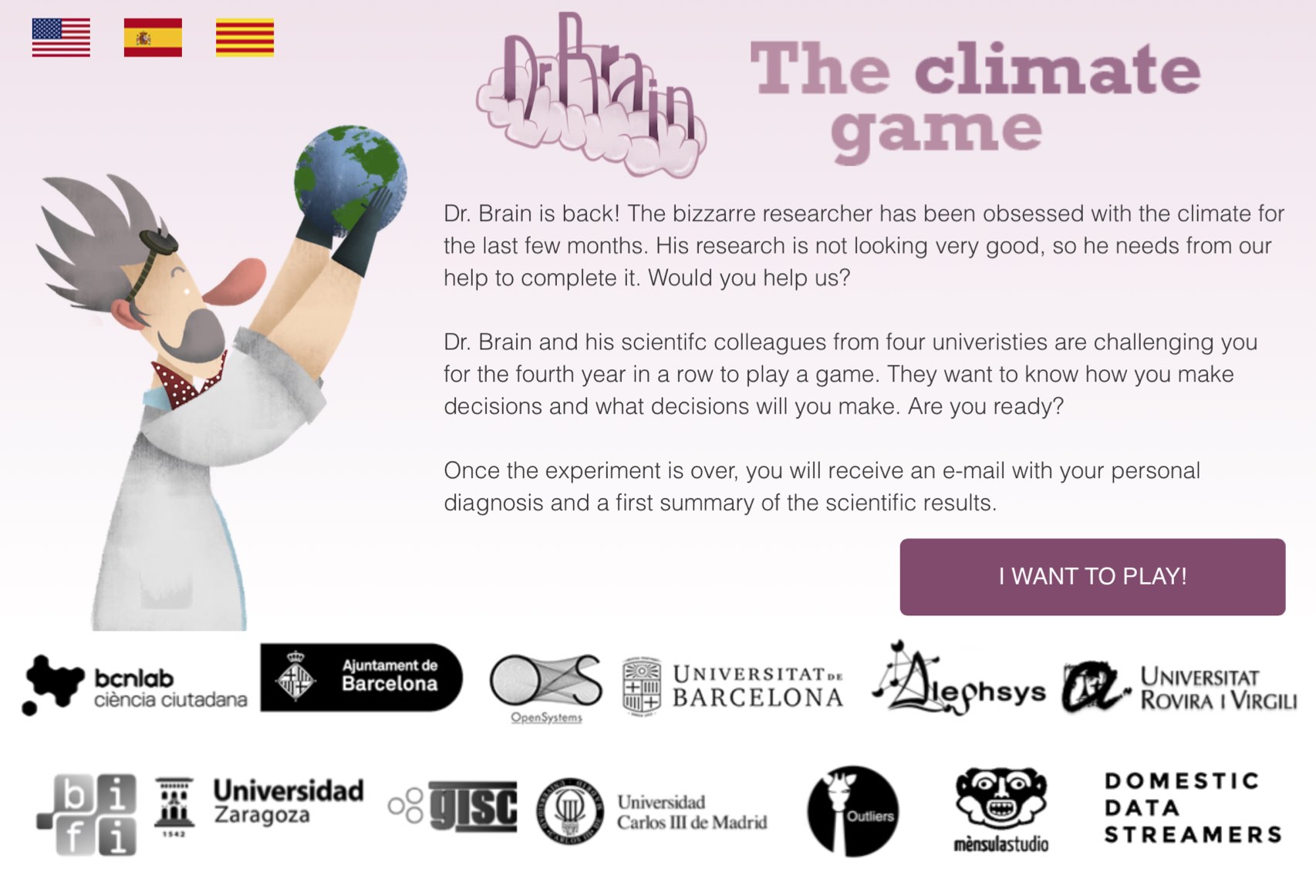}

\vspace{2mm} 

\end{array}$
\end{center}
\caption{Screenshots of the initial screen of three experiments (a) Mr. Banks, (b) Dr. Brain and (c) The Climate Game. In this screen we introduce a character and a narrative to attract the attention of the public and to motivate them to participate.}
\label{fig:initial}
\end{figure*}

\begin{figure*}[!th]  
\begin{center}$
\begin{array}{ll}
(a) \includegraphics[width=0.45\textwidth]{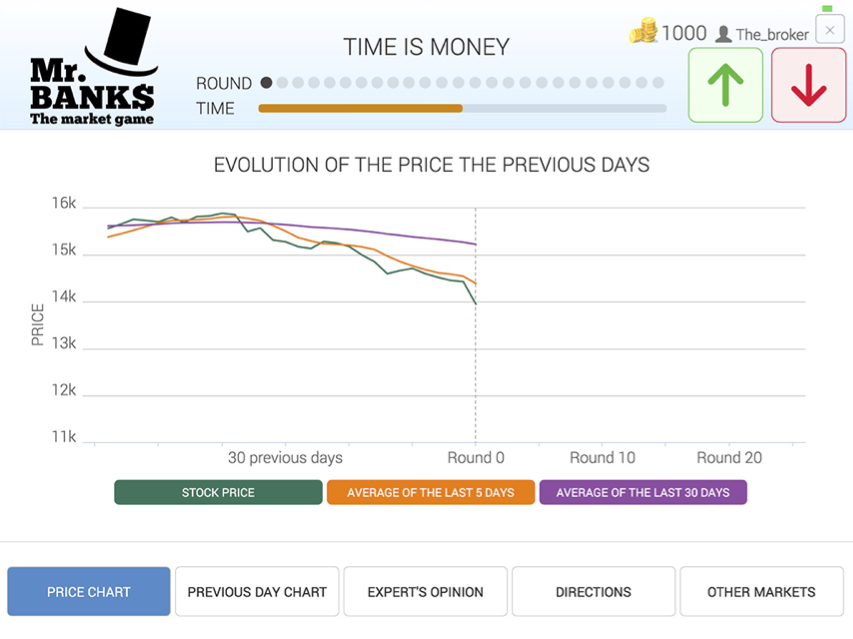}

\vspace{2mm} 

(b)
\includegraphics[width=0.45\textwidth]{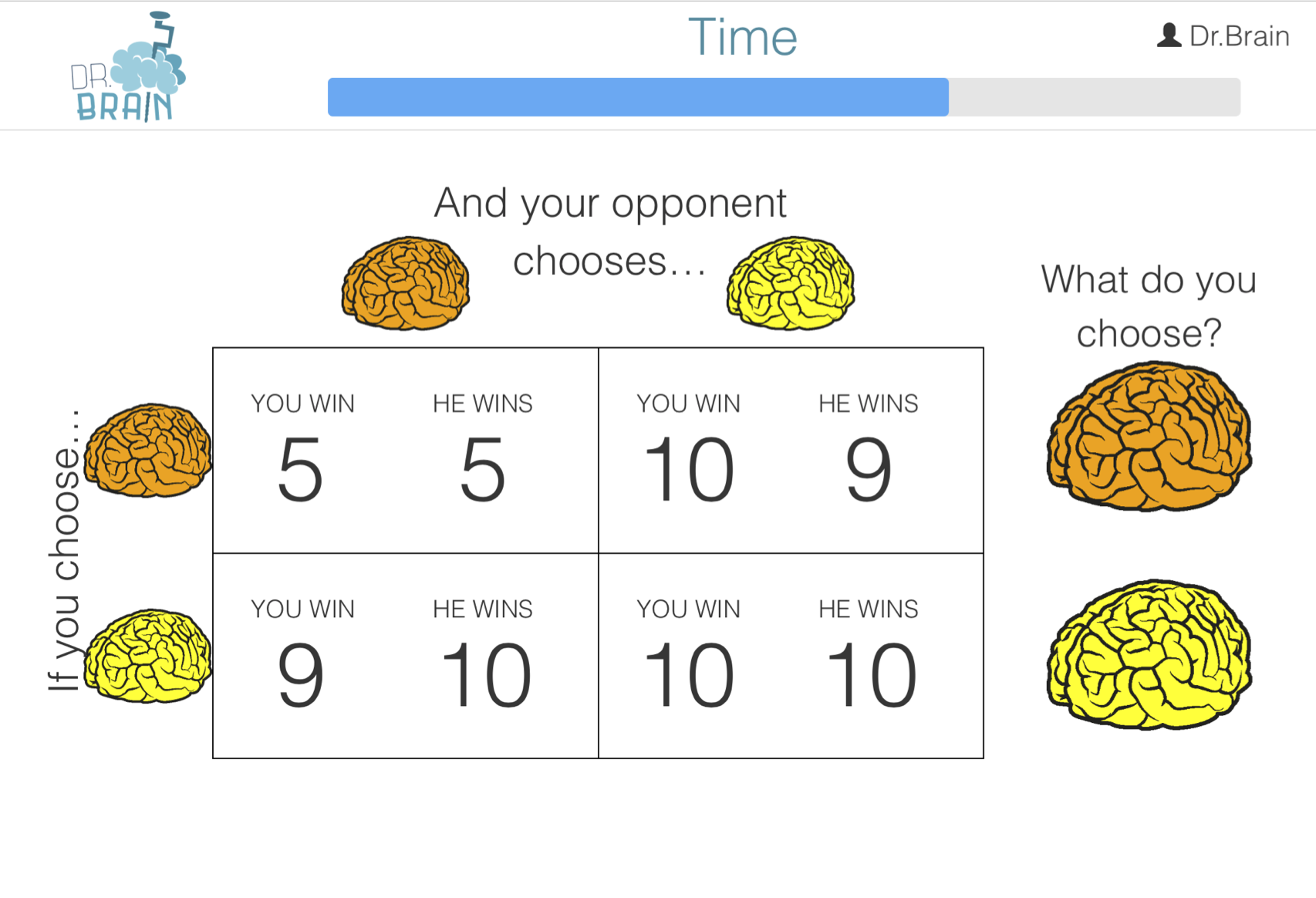}\\

\vspace{2mm}

(c) 
\includegraphics[width=0.45\textwidth]{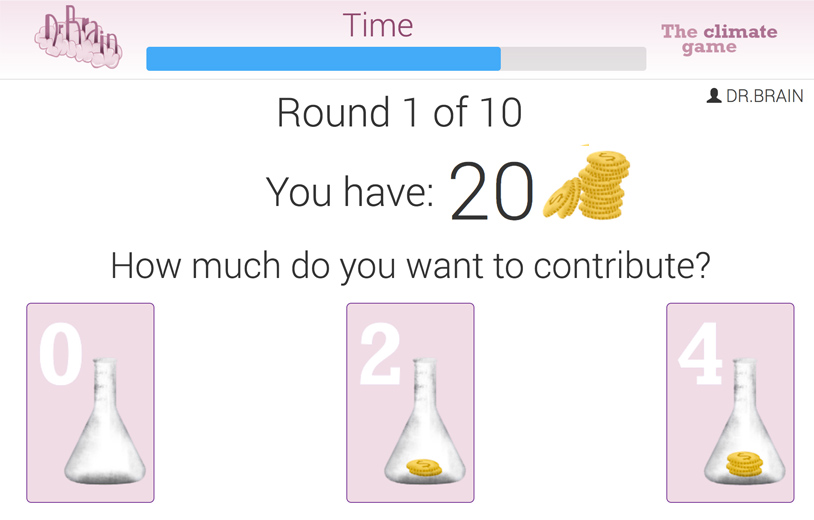}

\vspace{2mm}

\end{array}$
\end{center}
\caption{Screenshots of the main user interface of three experiments (a) Mr. Banks, (b) Dr. Brain and (c) The Climate Game where the participants respond to the dilemmas.}
\label{fig:game}
\end{figure*}

\begin{figure*}[!th]  
\begin{center}$
\begin{array}{ll}
(a) \includegraphics[width=0.45\textwidth]{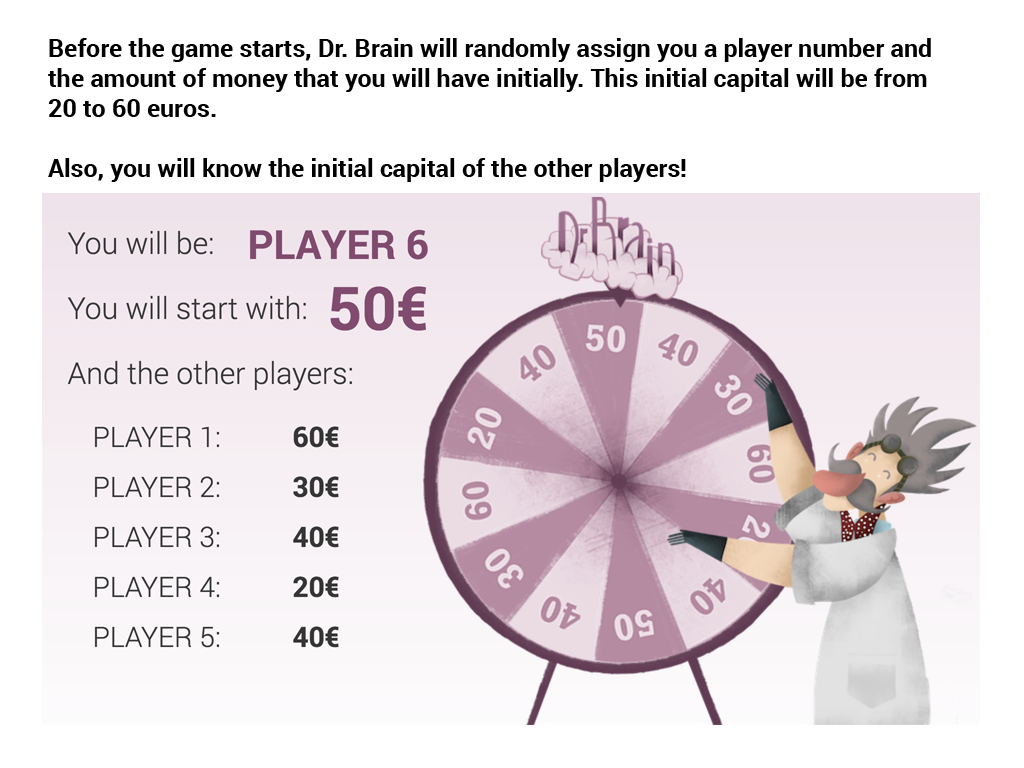}

\vspace{2mm} 

(b)
\includegraphics[width=0.45\textwidth]{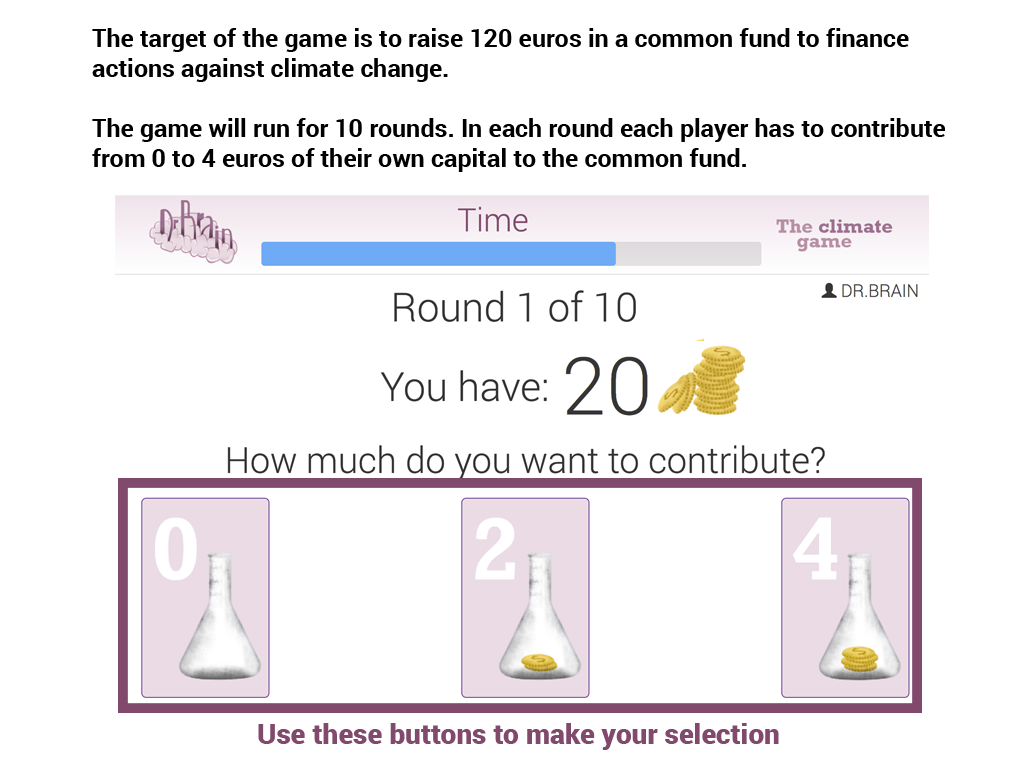}\\

\vspace{2mm}

(c) 
\includegraphics[width=0.45\textwidth]{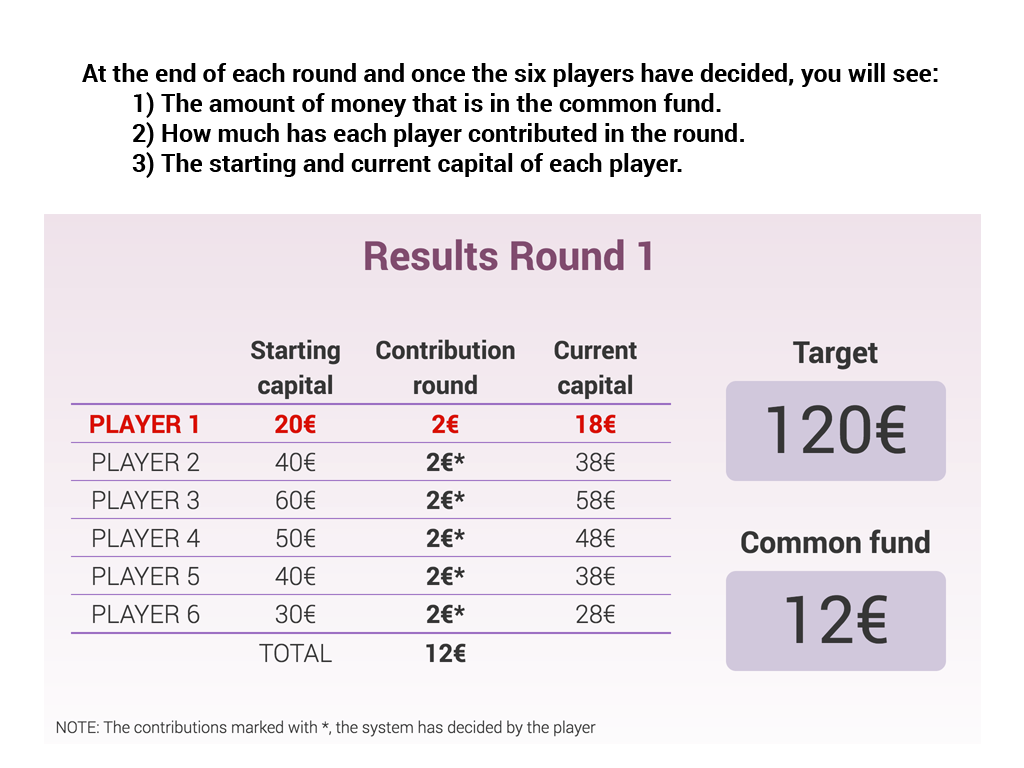}

\vspace{2mm} 

(d)
\includegraphics[width=0.45\textwidth]{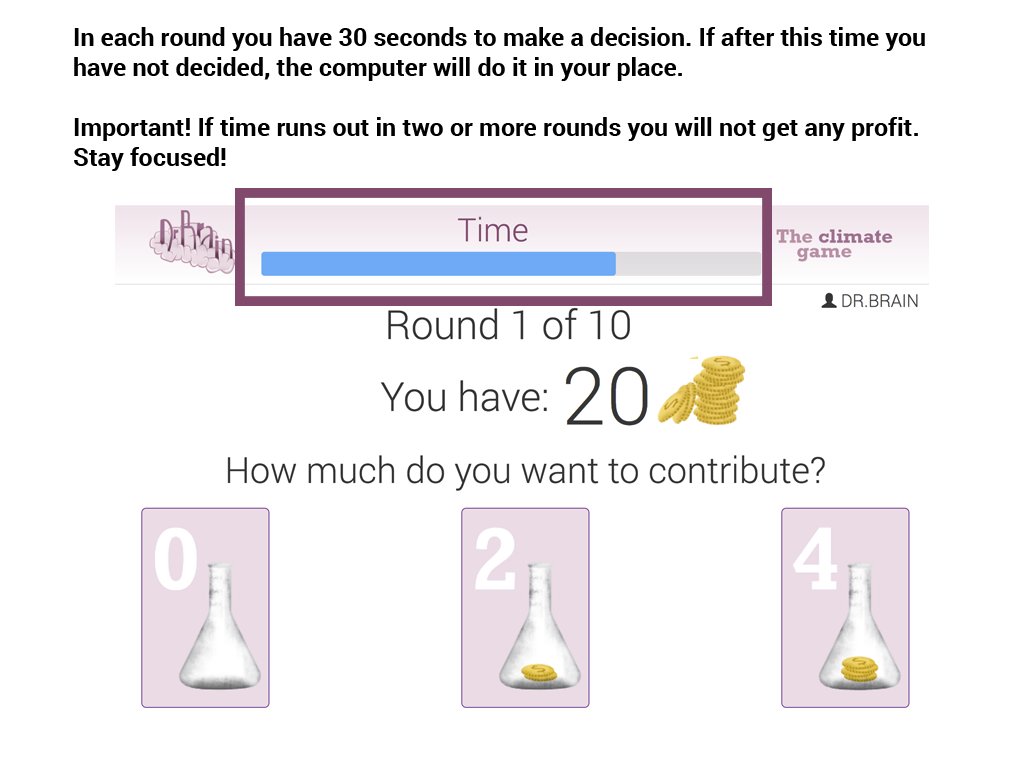}

\end{array}$
\end{center}
\caption{Screenshots of the tutorial shown before The Climate Game experiment where the participants learn the game mechanics and familiarize with the user interface.}
\label{fig:tutorial}
\end{figure*}

In this section we present some screenshots of the different experiments that have been deployed within the platform. First, in Fig. \ref{fig:initial} we present the initial screenshots of three of the experiments, Mr. Banks, Dr. Brain and The Climate Game. In these screens we observe the characters designed as part of the experimental setup to create a narrative and attract the attention of the public to the experiment. In Fig. \ref{fig:game} we show the user interface of these three experiments. In all the cases the interface uses a simplistic but visually appealing approach to present the dilemmas to the participants. Additionally, in all the experiments there is a tutorial where the participants learn the mechanics of the game and the user interface, we can see some of the screens of the tutorial of The Climate Game experiment in Fig.\ref{fig:tutorial}.

\clearpage
\newpage

\newpage
\section{Supplementary Tables}
\bigbreak

\begin{table}[h!]
  \centering
  \caption{Market imitation. Biases with respect to the market (Participant/Market)}~\label{tab:market_imitation}
  \begin{threeparttable}
  \begin{tabular}{C{3cm} C{2cm} C{2cm} C{2cm} c}
    \toprule
    {\small \textbf{Experiment}} 	& {\small \textbf{Up/Up}} & {\small \textbf{Up/Down}} & {\small \textbf{Down/Up}} & {\small \textbf{Down/Down}}\\
    \midrule
    DAU		&	0.71 			& 0.29		& 0.47	& 0.53\\	 
	CAPS	&	0.71 			& 0.29		& 0.49	& 0.51\\	
	Sonar+D	&	0.78$^\star$ 	& 0.22 		& 0.46	& 0.54\\	
	\bottomrule
  \end{tabular}
    \vspace{1ex}
	\raggedright $^\star$ There are significant differences (-2.53 SD) between DAU and Sonar+D experiments (Binomial process differences test).
  \end{threeparttable}
\end{table}

\begin{table}[h!]
  \centering
  \caption{Win-Stay Lose-Shift strategy. Decision conditioned to performance (Strategy/Decision)}~\label{tab:win_lose}
  \begin{threeparttable}
  \begin{tabular}{C{3cm} C{2cm} C{2cm} C{2cm} C{2cm}}
    \toprule
    {\small \textbf{Experiment}} 	& {\small \textbf{Win/Stay}} & {\small \textbf{Win/Shift}} & {\small \textbf{Lose/Stay}} & {\small \textbf{Lose/Shift}}\\
    \midrule
    DAU		&	0.68 	& 0.32		& 0.42	& 0.58\\	 
	CAPS	&	0.71 	& 0.29		& 0.48	& 0.52$^\star$\\	
	Sonar+D	&	0.72 	& 0.28 		& 0.40	& 0.60\\	
	\bottomrule
  \end{tabular}
  \vspace{1ex}
	\raggedright $^\star$ There are significant differences (2.35 SD) between DAU and CAPS experiments (Binomial process differences test).
  \end{threeparttable}
\end{table}

\begin{table}[h!]
\centering
\caption{Satisfaction of participants in Mr.Banks (n=234), Dr.Brain (n=524) and The Climate Game (n=420).}
\label{my-label}
\begin{tabular}{cccccc}
\toprule
 										{\small \textbf{Experiment}} & \textbf{Very Positive}	 & \textbf{Positive}  & \textbf{Neutral}  	& \textbf{Negative} 	& \textbf{Very Negative}  \\
 \midrule
 Mr.Banks						& - 				 & 125 		 & 91  		& 18			& - 			 \\
 Dr.Brain						& 245 			 & 217  		 & - 		& 49			& 13			 \\
 The Climate Change				& 204 			 & 184  		 & -		& 25			& 7				 \\
 \bottomrule
\end{tabular}
\end{table}

\clearpage
\newpage
\end{document}